# Helmet ultrasound for brain imaging in post-hemicraniectomy patients

Yang Zhang[1,2, †], Karteekeya Sastry[1,3, †], Iyla Rossi[1, †], Joshua Olick-Gibson[1], Jonathan J. Russin[4,5,6, *], Charles Y. Liu[4,5,6, *], Lihong V. Wang[1,3, *]

[1] Caltech Optical Imaging Laboratory, Andrew and Peggy Cherng Department of Medical Engineering, California Institute of Technology, 1200 East California Boulevard, Pasadena, CA 91125, USA.

[2] School of Biomedical Engineering, Tsinghua University, Beijing, 100084, China.

[3] Caltech Optical Imaging Laboratory, Department of Electrical Engineering, California Institute of Technology, 1200 East California Boulevard, Pasadena, CA 91125, USA.

[4] Department of Neurological Surgery, Keck School of Medicine, University of Southern California, Los Angeles, CA 90033, USA.

[5] Neurorestoration Center, Keck School of Medicine, University of Southern California, Los Angeles, CA 90033, USA.

[6] Rancho Los Amigos National Rehabilitation Center, Downey, CA 90242, USA.

[†]These authors contributed equally to this work: Yang Zhang, Karteekeya Sastry, Iyla Rossi

*Corresponding authors. J. J. Russin (jonathan.russin@med.usc.edu), C. Y. Liu (cliu@usc.edu), L. V. Wang (LVW@caltech.edu)

## Abstract

Noninvasive imaging deep into the adult brain at submillimeter and millisecond scales remains a challenge in medical imaging. Here, we report a helmet-based ultrasound brain imager—built from a customized helmet, a scanned ultrasound array, and three-dimensional (3D) printing—for real-time imaging of human brain anatomical and functional information. Through its application to post-hemicraniectomy patients in a sitting position, we achieved 3D brain tissue structural, vascular, and blood flow images at centimeter-scale depths with submillimeter and millisecond spatiotemporal resolutions. We also demonstrated the system's capability to track cerebral blood flow over repeated imaging sessions, including during motion-prone conditions. Our brain imager circumvents the skull and bridges the gap between high-resolution human brain imaging and wearable convenience. This imager may serve as a platform for further investigations into human brain dynamics in post-hemicraniectomy patients and offer insights into the brain that could surpass those obtained from non-human primate studies.

**Introduction**

Human brain imaging has traditionally relied on established modalities, such as magnetic resonance imaging (MRI)[1], x-ray computed tomography (CT)[2], and positron emission tomography (PET)[3], each with its unique set of strengths and limitations. Imaging with MRI provides detailed brain anatomy and functional MRI (fMRI)[1] captures blood oxygen level-dependent (BOLD) signals associated with brain activity to map clinically relevant function to detailed anatomical images. Nevertheless, fMRI exhibits a non-linear relationship with hemoglobin concentration, is extremely costly, and is unable to be adapted for wearability. CT scans are relatively cheaper compared with MRI, are rapid to acquire and can be portable. However, CT utilizes ionizing radiation, which significantly limits its application, and is not able to image brain function. PET scans, on the other hand, can be utilized to image brain function, however they require the administration of radioactive tracers for metabolic activity measurements.

While traditional methods have significantly contributed to our understanding of the brain, the pursuit of wearable brain imaging technologies has gained momentum. In response to the need for portable and adaptable brain imaging solutions, electroencephalography (EEG)[4] and functional near-infrared spectroscopy (fNIRS)[5] have been explored. EEG, a wearable brain monitoring device, offers high temporal resolution but lacks spatial precision. Similarly, fNIRS, another wearable technology for monitoring brain function, faces spatial resolution limitations. There remains a compelling need for the development of a wearable brain imaging device with both high temporal and spatial resolutions, to address the current constraints in brain imaging technologies.[6]

Ultrasound imaging has evolved into an invaluable healthcare tool, offering a safe and versatile means of visualizing various organs and tissues within the human body[7]. Its applications include monitoring fetal development[8], evaluating muscular system diseases[9,10], assessing heart function[11,12], and examining the abdomen[13,14] and brain[15–17]. One notable domain of ultrasound imaging is human brain ultrasound, specifically transcranial ultrasound, such as transcranial

Doppler ultrasound (TCD)[18]. TCD is employed to assess cerebral blood flow in major cerebral arteries, aiding in the diagnosis of conditions such as stroke, arterial spasm and the evaluation of cerebral disorders[19]. Another significant application of ultrasound is fetal brain imaging[20]. It plays a pivotal role in monitoring conditions such as fetal brain lesions[21] and intraventricular hemorrhage[22]. Recent advancements have demonstrated the potential of functional ultrasound in imaging brain activity in human newborns[15] and providing bedside functional monitoring of dynamic brain connectivity in neonates[23]. Furthermore, functional ultrasound has shown clinical promise in intraoperative functional and vascular brain mapping during awake brain surgery[24], and it also achieves microscopic transcranial imaging of human cerebral vasculature through the use of contrast agents[25].

Recently, the innovative concept of wearable ultrasound has emerged in the field of human body imaging. Notably, a bioadhesive ultrasound (BAUS)[26] device has been engineered to facilitate prolonged imaging, extending up to 48 hours, of various internal organs and critical structures such as blood vessels, muscles, the heart, and the lungs. Additionally, an innovative wearable ultrasound cardiac imager has been devised to enable continuous, real-time, and direct assessment of cardiac function[27]. This technology boasts significantly enhanced accuracy across diverse environmental conditions. A conformable ultrasound patch has also been developed to enable deep tissue scanning and imaging over large-area curvilinear organs, such as the breast[28]. A low-cost wearable ultrasound device was also developed for wrist and hand kinematic tracking, serving as a human-machine interface[29]. Recently, a conformal ultrasound patch was developed for transcranial imaging of blood flow in major cerebral arteries, enhancing the practicality of conventional TCD by offering a hands-free, non-invasive solution[30]. Despite these remarkable strides, one notable gap remains: the absence of a wearable ultrasound design specifically tailored for human brain imaging with submillimeter and millisecond spatiotemporal resolutions.

This article explores the development and capabilities of a helmet-based brain ultrasound imager designed to bridge the gap between high-resolution brain imaging and portable, user-friendly technology. We present a helmet-based ultrasound scanner capable of deep brain imaging with submillimeter spatial and millisecond temporal resolution. Optimized for post-hemicraniectomy patients, the system enables structural and functional brain imaging in a wearable setting. Using ultrasound, this approach facilitates real-time structural and blood flow imaging, providing a practical tool for neuroimaging research. Phantom validation demonstrates the system's ability to generate three-dimensional (3D) ultrasound images of structural and blood flow patterns in line matrix targets, *ex vivo* blood flow phantoms, and *ex vivo* porcine muscle. *In vivo* experiments in post-hemicraniectomy patients further show that the system can capture brain tissue structures at centimeter depths, including cortical cerebral blood flow. Additionally, the system enables repeated blood flow imaging over repeated imaging sessions. These results establish our helmet-based wearable ultrasound as a promising platform for accessible brain imaging and longitudinal neuroimaging studies in post-hemicraniectomy patients.

**Results**

The helmet-based ultrasound brain imager, as shown in Fig. 1a and Supplementary Fig. 1, comprises four integrated components: an ultrasound array for transmitting ultrasound pulses and detecting backscattered ultrasound waves from the head, a motorized linear translation stage for scanning the array to form a 3D field-of-view, a customized helmet to support the ultrasound array with the motor and ergonomically fit on the human head, and custom-made 3D-printed parts to provide flexibility in adjusting the position and orientation of the ultrasound array. Our helmet-based design not only ensures the best fit for the imaging probe on the human head but also provides a window for coupling the ultrasound array with the scalp. We have the flexibility to select the ultrasound array imaging sequence, which can generate either a focused beam or a tilted plane wave (refer to Supplementary Fig. 2), thus allowing us to obtain ultrasound images of the human brain and the scalp in real-time mode or ultrafast mode, respectively. The addition of a motorized linear translation stage aids in the identification and adjustment of the imaging position for two-dimensional (2D) slice imaging and facilitates 3D imaging through linear scanning of the ultrasound array. Furthermore, the 3D-printed connecting joints offer freedom of movement in multiple directions (refer to Supplementary Fig. 3), ensuring that the ultrasound array maintains acoustic coupling with the scalp and acquires data from the areas of interest.

The front view of a schematic of the system is shown in Fig. 1b, illustrating the attachment of the ultrasound array to the motor, which is in turn mounted onto the helmet. In Fig. 1c, a side view of the system schematic highlights the motor's role in enabling the acquisition of 3D volumes of the target. The combination of the motor and the 3D-printed parts collectively provides the essential flexibility to position the ultrasound array over various regions of the head and at different angles, thus accommodating variations in head shape. The helmet ultrasound imager has dimensions of approximately 33 cm (length) × 28 cm (width) × 23 cm (height) in the $x$, $y$, and $z$ directions, with a total weight of 1.3 kg. For this study, we consider subjects who have undergone a hemicraniectomy with intact scalp. We imaged the brains of four post-hemicraniectomy subjects (3 males and 1 female, two left-sided and two right-sided craniectomy) using our helmet-based wearable ultrasound imager in a sitting posture. An example ultrasound brain recording from one of the subjects is shown in Fig. 1d. The acquired ultrasound signals reveal rich morphological and functional information about the human brain. Firstly, they are used to reconstruct structural ultrasound images of the brain, as shown in Fig. 1e. Secondly, power Doppler processing is performed on the ultrasound signals to reconstruct functional ultrasound (fUS) images of cerebral blood flow, as shown in Fig. 1f. Finally, the signals are utilized to track cerebral blood flow in motion-prone scenarios, as shown in Fig. 1g.

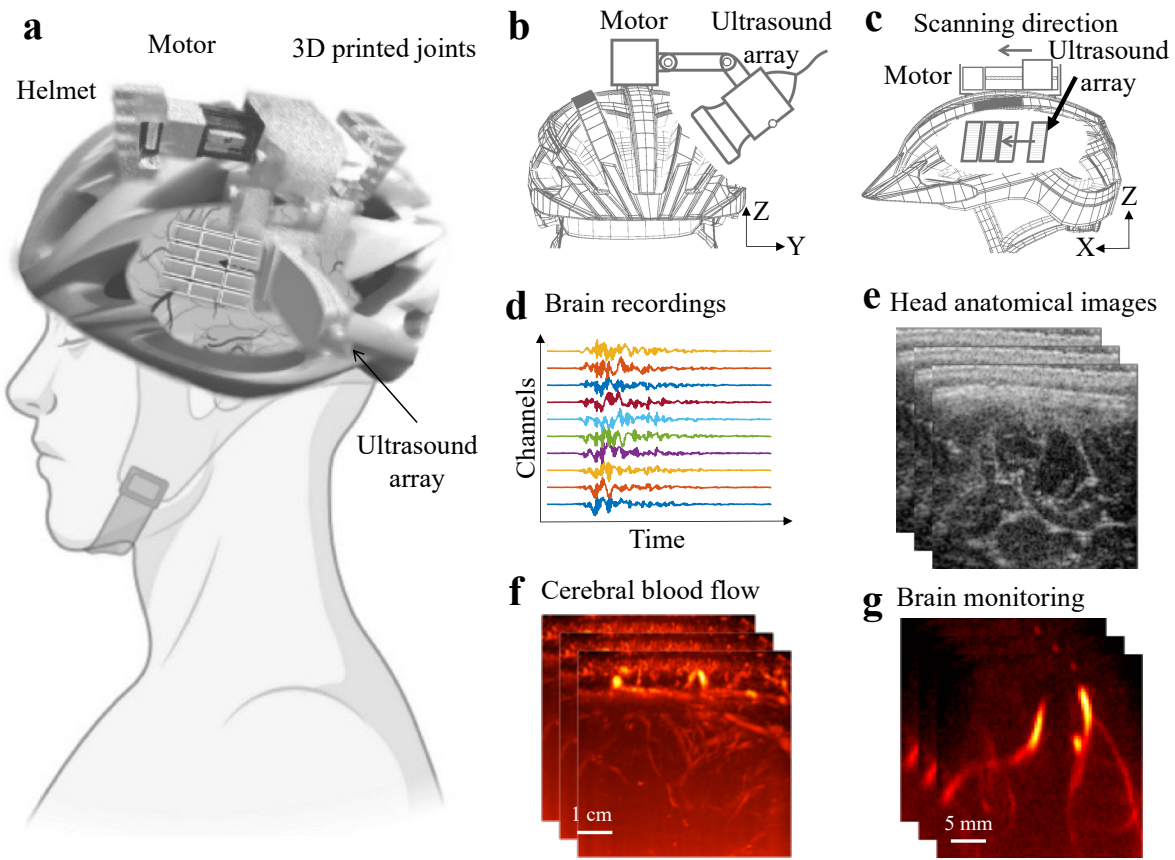

**Fig. 1 | Helmet ultrasound for brain imaging. a**, Schematic of the helmet-mounted ultrasound imaging probe. It comprises four major integrated components: 1) the customized helmet; 2) the ultrasound transducer array; 3) the motorized linear translation stage; 4) 3D-printed joints. **b,** Front view of a schematic of the wearable imaging device. The 3D printed joints ensure that the ultrasound array can be adjusted to conform to the surface of the head. **c**, Side view of a schematic of the wearable imaging device. Linear scanning of the motor enables the acquisition of three-dimensional images using a synthetic two-dimensional (2D) array. **d**, Representative raw ultrasound signals from multiple array elements (channels). **e**, Representative anatomical head images obtained using the device. **f**, Representative cerebral blood flow images. **g**, Monitoring of cerebral blood flow in motion-prone scenarios.

To characterize the ultrasound imaging performance of the system, we imaged a 3D line-target phantom comprising parallel wires (diameter: 150 $\mu$m) arranged in a 4×3 grid with a pitch of 1 cm, as illustrated in Supplementary Fig. 4a. The reconstructed volume of the line-target phantom is shown as maximum amplitude projections (MAPs) in two orientations (shown in the bottom-left corner of each image) in Figs. 2a and 2b, respectively. We also estimate the axial and lateral resolution of the imaging system using the 2D slice shown in Fig. 2c to be 400 $\mu$m and 500 $\mu$m, respectively. The elevational resolution was determined to be 1 mm by transversely scanning the line-target phantom. To demonstrate the fUS performance of the system, we imaged a 3D *ex vivo* blood flow phantom comprising arbitrarily arranged tubes (MRE080-S3850, Braintree scientific; outer diameter: 2.03 mm, inner diameter: 0.97 mm) with blood, meant to mimic blood flow within the human brain, as shown in Supplementary Fig. 4b. We extract the power Doppler signals for each 2D slice acquired at a frame rate of 500 Hz and plot the maximum projections in two orientations (shown in the bottom-left corner of each image) of the resulting 3D power Doppler volume in Figs. 2d and 2e. We also show one of the 2D slices in Fig. 2f. These images demonstrate the capability of our system to obtain 3D volumetric maps of blood flow. Finally, we imaged *ex vivo* porcine muscle, meant to mimic the human brain tissue, as shown in Figs. 2g and 2h. Both images show a clear separation between the skin and the muscle fibers. These *ex vivo* demonstrations illustrate our system's ability to image tissue structure, as well as blood flow in the brain of hemicraniectomy subjects.

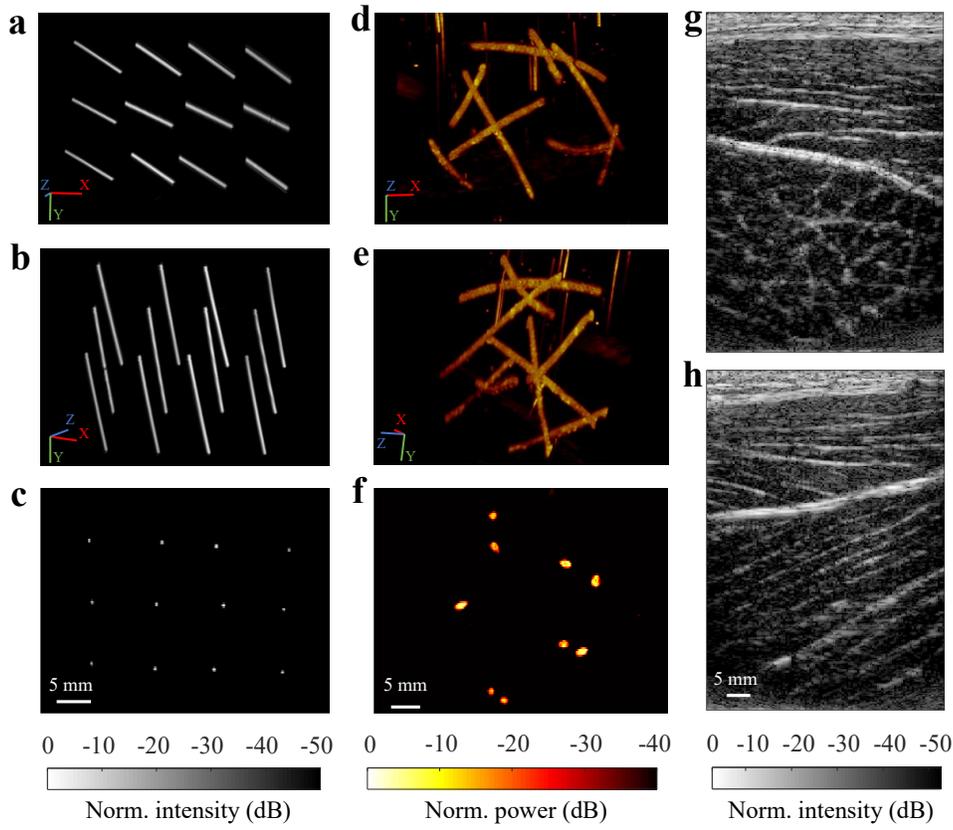

**Fig. 2 System characterization and phantom validation. a-b**, Maximum amplitude projections (MAPs) of the 3D line-target phantom in two orientations (shown in the bottom-left corner of each image) obtained using ultrasound. The line-target phantom comprises parallel wires (diameter: 150 $\mu$m) arranged in a 4×3 grid. **c**, 2D slice of the ultrasound image of the line-target phantom. **d-e**, Maximum projections in two orientations (shown in the bottom-left corner of each image) of the 3D volumetric blood flow map of a blood flow phantom obtained using fUS. The blood phantom comprises tubing (MRE080-S3850, Braintree scientific; outer diameter: 2.03 mm, inner diameter: 0.97 mm) arranged in arbitrary directions with circulating blood. **f**, 2D slice of the fUS image of the blood flow phantom. **g-h**, 2D slices of the ultrasound images of porcine muscle showing a clear separation between the skin and the muscle fibers.

Representative anatomical brain images from one of the hemicraniectomy patients (male, left-sided craniectomy), acquired using wearable ultrasound, are presented in Fig. 3. Fig. 3a displays brain structural images in a coronal view. Notably, it captures a major vessel branch with several sub-branches (indicated by white arrows) at depths ranging from 2 cm to 4 cm. These vessels are distinguishable from the surrounding tissue due to stronger backscattered signals from the vessel walls. Fig. 3b showcases two ventricles in the deep brain region, located at depths of 3.5 cm and 5 cm, respectively (indicated by white arrows). The four adjacent positions illustrate slight changes in the structure of the ventricular slices. The ventricles appear darker than the surrounding tissues due to weaker backscattered signals. The contrast-to-noise ratios (CNRs) of the two ventricles, computed as the average of the CNRs at 10 locations on the circumferences of the respective ventricles, are 17.2 and 14.8, respectively. In Fig. 3c, structural images from a different position on the hemicraniectomy side are displayed, clearly delineating the boundaries between the scalp (orange arrow), grey matter (dark arrow), white matter (white arrow), and deep brain features (green arrow), consistent with those in Fig. 3a and Fig. 3b.

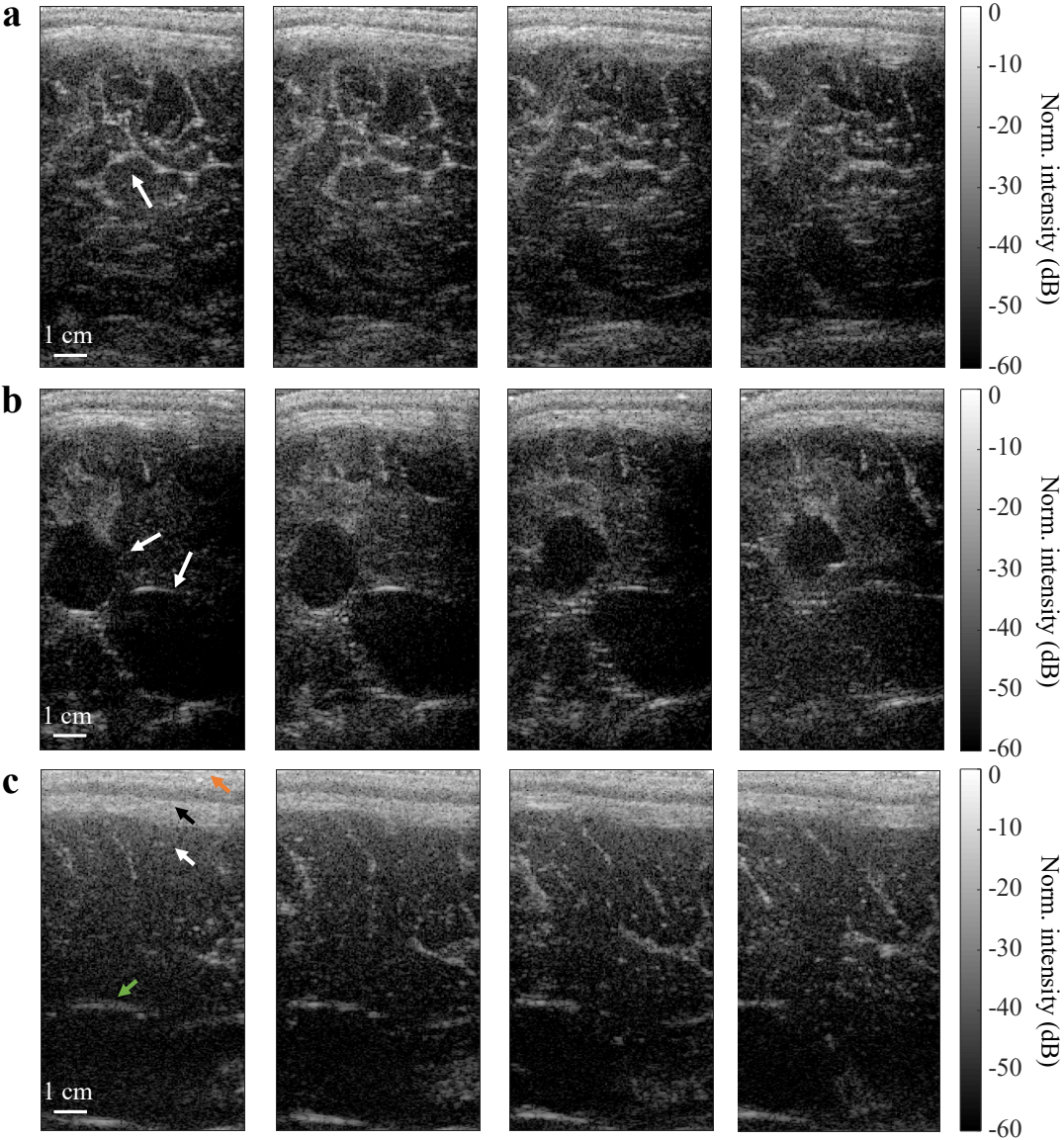

**Fig. 3 | *In vivo* human brain anatomical images. a**, Cross-sectional brain structural images of a hemicraniectomy patient showing vessel branches (white arrow) at four adjacent positions with a pitch of 2 mm. **b**, Cross-sectional brain structural images showing ventricles (two white arrows) at four adjacent positions. **c**, Cross-sectional structural images of the brain showing scalp (orange arrow), brain grey matter (dark arrow), white matter (white arrow), and deep features (green arrow).

Representative cerebral blood flow images obtained through power Doppler processing from three hemicraniectomy patients (2 males and 1 female, one left-sided and two right-sided craniectomy) are presented in Fig. 4. Fig. 4a displays 3D-projected blood flow images of the hemicraniectomy side in a coronal view. Delineated by the dashed line, both blood vessels in the scalp region and the brain region are discernible. The vasculature network, featuring multiple branches, is clearly resolved in the brain cortex region. Figs. 4b-d exhibit three example 2D cross-sectional blood flow images of the brain, each illustrating distinct vasculature patterns. Figs. 4e-g comprise representative 3D-projected blood flow images from the three patients, encompassing a left-side hemicraniectomy male, a right-side hemicraniectomy male, and a right-side hemicraniectomy female subject, respectively. The average CNRs (mean ± standard error) of the vascular structures corresponding to three hemicraniectomy subjects in Figs. 4e-g are $20.4 \pm 3.3$, $16.9 \pm 2.9$, and $24.2 \pm 3.9$, respectively. We also quantify the diameters of some of the fine vessels in a representative 2D blood flow image to be close to 0.5 mm in Supplementary Fig. 5. Together, these brain vasculature and blood flow images demonstrate the feasibility of our wearable ultrasound for imaging cerebral blood flow.

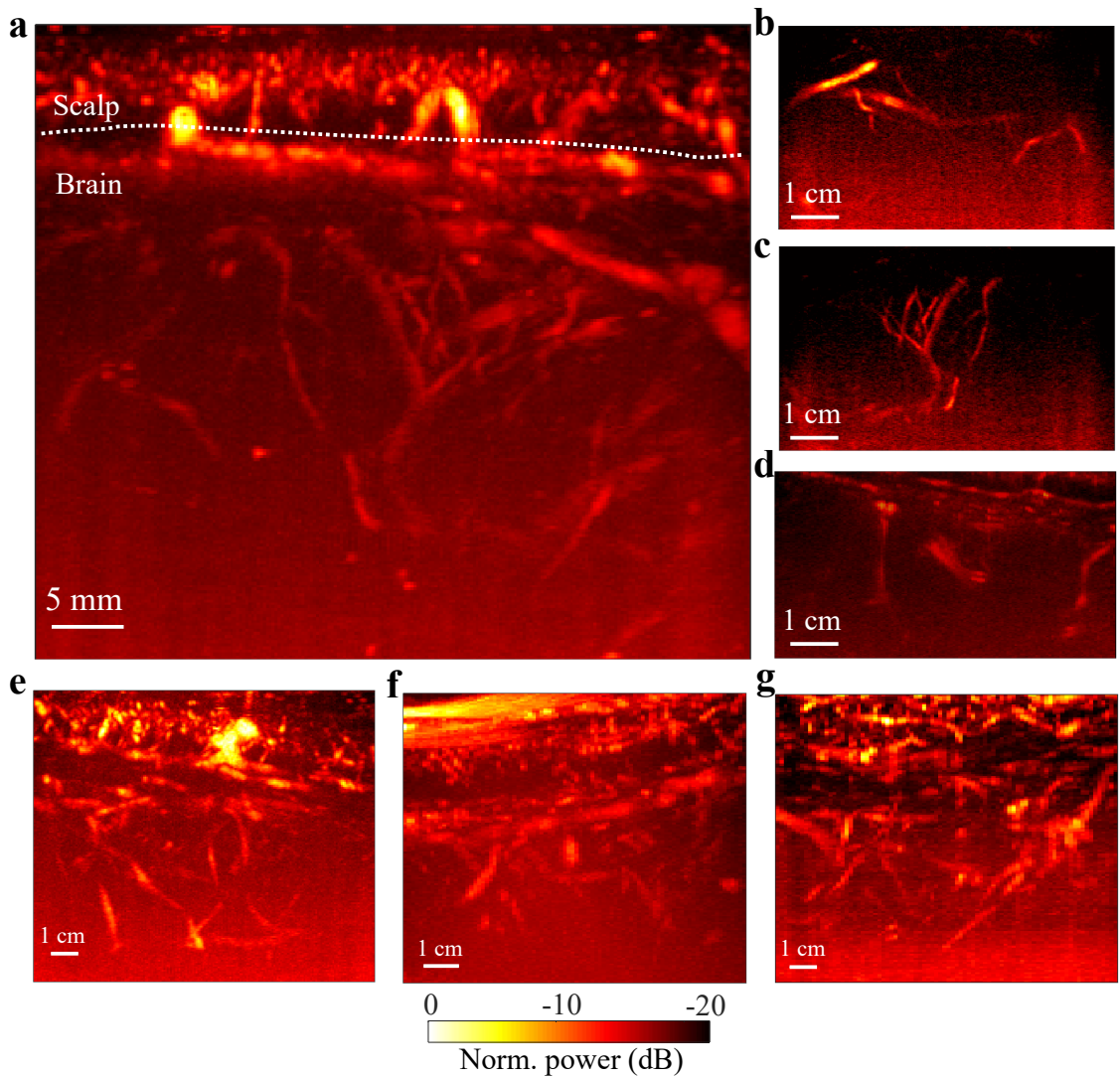

**Fig. 4 | *In vivo* human brain blood flow images. a**, Representative maximum projection of the 3D blood flow image of a hemicraniectomy patient, illustrating the vasculature network in the scalp and brain regions. **b-d**, Cross-sectional 2D blood flow images displaying vasculature branches in three positions of the hemicraniectomy side in a coronal view. **e-g**, Maximum projections of 3D blood flow images from the three patients, encompassing a left-side hemicraniectomy male, a right-side hemicraniectomy male, and a right-side hemicraniectomy female subject, respectively.

Lastly, we acquired blood flow images of the brain of a hemicraniectomy subject (male, left-sided craniectomy) seated in a chair over a period of 120 seconds with a frame rate of 0.4 Hz. The subject was prone to involuntary head and body movements, resulting in significant motion during the acquisition, as illustrated schematically in Fig. 5a. This can be observed from the averaged blood flow images (over 120 seconds) at two different positions shown in Figs. 5b and 5c on the left, which are noticeably blurred. However, this inter-frame motion can be corrected (see Methods), as shown on the right in Figs. 5b and 5c. The peak CNRs of the images in Figs. 5b and 5c are improved by 86% and 34%, respectively, after motion correction. We also display the individual frames acquired at the two positions at 0, 40, 80, and 120 seconds before and after motion correction in Figs. 5d-g. From these images, we can see that despite significant motion, we can visualize and track the cortical vessel structures in each position over the entire duration.

In a stable setup with minimal motion, we imaged a subject (male, left-sided craniectomy) over a 10-minute period at a frame rate of 0.4 Hz, observing stable cerebral blood flow during these extended acquisitions. It takes less than 1 minute to wear the helmet ultrasound imager. Additionally, we demonstrate repeated brain vasculature imaging in Supplementary Fig. 6, showcasing two blood flow images of the same target within the brain of a hemicraniectomy subject (in a sitting position), captured 137 minutes apart. After acquiring the first image, the helmet was removed and then re-secured on the subject's head two hours later. Despite the interval and repositioning, we observed similar brain vasculature features in the two images, with minor discrepancies attributed to a slight shift in the probe's imaging position. This demonstrates our device's ability to perform repeated cerebral blood flow imaging, even in non-recumbent, motion-prone scenarios, highlighting its potential as a portable and user-friendly brain imaging tool.

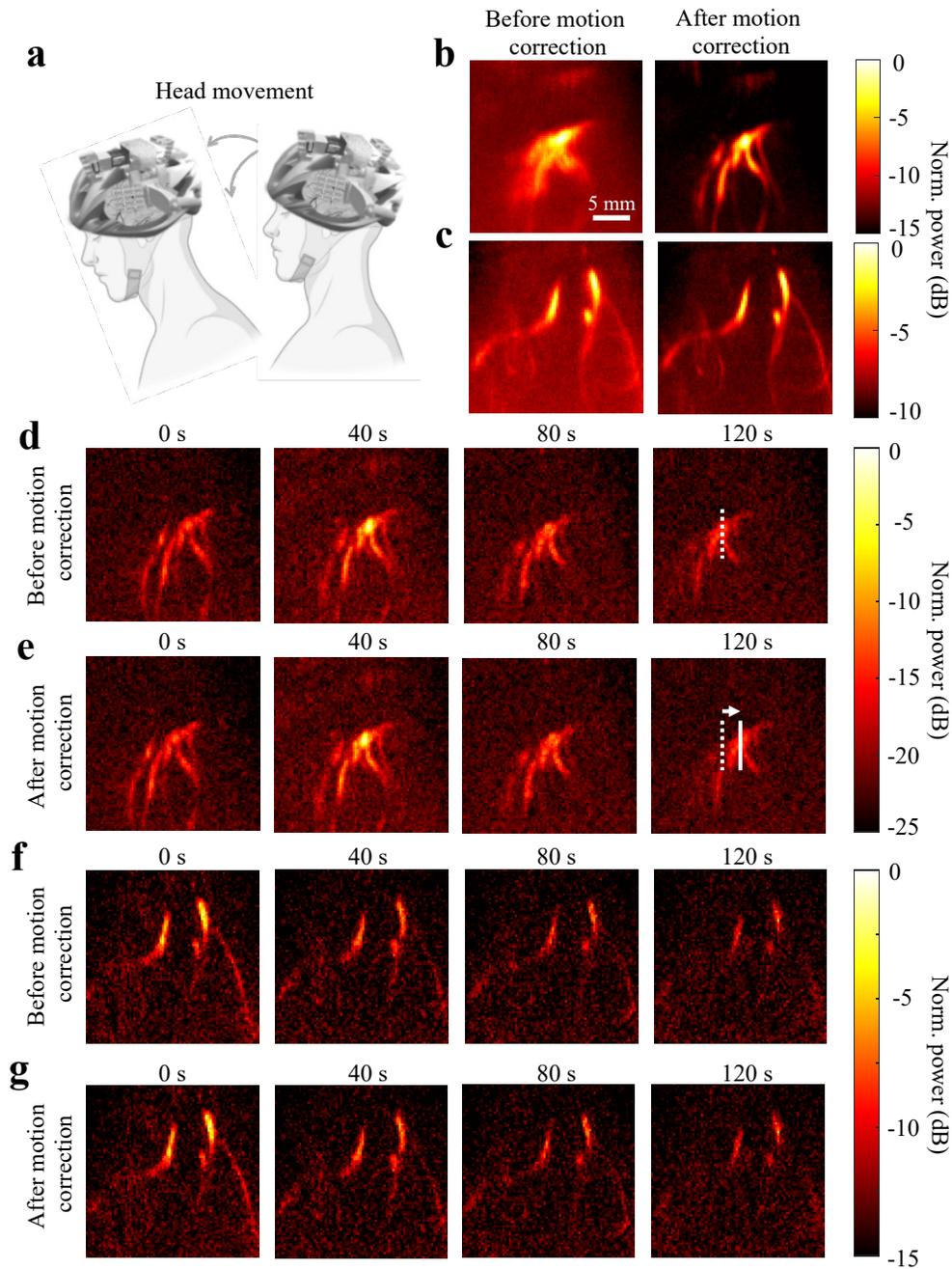

**Fig. 5 | Monitoring of blood flow in the brain in a non-recumbent motion-prone situation. a,** Schematic of the human brain monitoring experiment. We acquired data from a hemicraniectomy subject in a seated position over a 120-second period with a frame rate of 0.4 Hz. Despite significant involuntary head and body motion during the acquisition, we were able to monitor blood flow in the brain during this period. **b-c,** Averaged blood flow images over 120 seconds at two different positions before (left) and after (right) motion correction. **d**, 2D blood flow images at location #1 at 0, 40, 80, and 120 seconds before motion correction. **e**, 2D blood flow images at location #1 at 0, 40, 80, and 120 seconds after

motion correction. Dashed and solid white lines in the rightmost panels of **d** and **e**, respectively, illustrate the extent of the motion. **f**, 2D blood flow images at location #2 at 0, 40, 80, and 120 seconds before motion correction. **g**, 2D blood flow images at location #2 at 0, 40, 80, and 120 seconds after motion correction.

## Discussion

This study introduces the concept of a helmet-based wearable ultrasound for imaging the human brain, showcasing its potential to capture both anatomical and functional information. We characterize the system's performance in 2D and 3D structural and blood flow imaging through phantom experiments. Through *in vivo* brain imaging of post-hemicraniectomy patients, we demonstrate the feasibility of wearable ultrasound for imaging brain tissue structures at depths of several centimeters, allowing for clear differentiation between the scalp layer, grey matter, white matter, and deep tissue features. Additionally, we show that the imager can measure blood flow using power Doppler processing, particularly in the cortical region. This technology provides insights into vascular anatomy within both the scalp and brain in post-hemicraniectomy patients. In the context of mapping brain function, we demonstrated the system's ability to track cerebral blood flow across repeated imaging sessions, including during motion-prone conditions. Although the full volumetric scan requires more time than a single 2D slice acquisition, the system's mechanically stable helmet design ensures consistent transducer-to-scalp coupling, enabling high-fidelity data acquisition even in the presence of minor subject movements. The imager circumvents the skull, providing a new approach to high-resolution brain imaging with wearable convenience. This system holds promise for advancing real-time, non-invasive brain imaging in post-hemicraniectomy patients and could serve as a platform for future investigations into human brain dynamics[31], bridging gaps in understanding that have traditionally been addressed through non-human primate studies.

Currently, we have customized a helmet to incorporate an acoustic window and mounts for the motorized linear translation stage. Alternatively, we can design and manufacture a helmet from scratch, using techniques such as 3D printing. As hemicraniectomy patients may undergo craniectomy on either side of the brain, we have developed two versions of the helmet, one for the left side and one for the right side of the brain imaging window. Furthermore, we prepared two sizes of the customized helmet with adjustable straps to accommodate patients of varying head sizes. For the ultrasound probe, we have connected a customized probe to a motorized linear translation stage with 3D-printed joints, allowing for flexible adjustment of the probe's position and orientation for imaging different regions of the human brain. The motorized linear translation stage also enables our system to acquire 3D volumes by stitching together 2D slices. One potential improvement could be made by utilizing a 2D matrix array probe to achieve single-shot 3D imaging. This can be easily incorporated into our design by modifying the probe holder (see Supplementary Fig. 3). Currently, our wearable ultrasound imager is based on a customized helmet. However, we are exploring the possibility of creating a more integrated version, such as the development of a wearable ultrasound patch, which would be significantly smaller and lighter. In this study, we chose to implement a 128-element, 5 MHz linear array transducer as a tradeoff between imaging depth, resolution, cost, computation load, and system complexity. While commercial probes with higher element counts are available and may offer improved image quality, our primary goal was to establish a proof-of-concept for the wearable helmet-based imaging system and validate its performance under practical conditions. The current design prioritizes adaptability, and the helmet is readily compatible with other array configurations, including 256-element or higher-density probes. Future work will explore integrating such transducers to optimize image quality and broaden clinical applicability, leveraging the modular nature of our helmet platform.

In the current prototype, helmet placement is quick (under one minute), but slight variations in helmet positioning or transducer alignment across sessions can affect the reproducibility of image acquisition. To address this in future studies, we plan to implement several improvements. First, we are designing mechanical alignment guides (e.g., adjustable brackets or positioning pegs) that lock the helmet into a consistent orientation relative to the subject's anatomical landmarks. Second, we are developing a fiducial-based calibration system, in which embedded markers on the helmet or transducer surface can be co-registered with anatomical features through either optical tracking or initial ultrasound scans. This will enable automated correction for positional drift and facilitate reproducible multi-session imaging. Third, we are exploring the integration of 3D-printed subject-specific head molds or pads, which conform to the individual's scalp contour and ensure consistent contact and orientation.

Here, we demonstrated the feasibility of using helmet-based wearable ultrasound for brain imaging in hemicraniectomy patients, including 3 males and 1 female, with cases involving both left- and right-side hemicraniectomies. We were able to obtain high-quality images depicting brain tissue anatomy, vascular structures, and cerebral blood flow in the cortex vessels. This is attributed to the deep penetration ability of ultrasound, its Doppler effect on moving blood scatterers (at a center frequency of a few MHz), and the absence of a skull barrier. We currently perform power Doppler processing to obtain blood flow instead of color Doppler processing due to its higher sensitivity for detecting vascular blood flow, as demonstrated in Supplementary Fig. 7. Additionally, the acoustic aberration induced by the human skull poses a challenge to transcranial ultrasound imaging. Due to the significantly different acoustic properties of the skull relative to soft tissue, acoustic waves at the skull-tissue interfaces undergo reflection and mode-conversion into compression and shear waves[32,33]. The situation is exacerbated by the dispersion and frequency-dependent attenuation induced by the skull. While our current imaging implementation does not yet achieve high spatial and temporal resolution through the intact skull, our study involving a unique population of post-hemicraniectomy patients presents a valuable opportunity for further investigations into human brain dynamics and offers insights that could surpass those obtained from non-human primate studies. Despite this, the wearable ultrasound concept has the potential to be adapted for transcranial imaging, combining advanced de-aberration techniques[34] to image adult brain structures and functions with high spatiotemporal resolution.

In our experiments with post-hemicraniectomy patients, we positioned the subjects in a sitting posture to demonstrate the blood flow tracking capability of our device. We chose this position because patients who have undergone hemicraniectomy may experience reduced mobility, making prolonged standing more challenging. Additionally, the sitting position offers greater convenience for conducting various brain function studies compared to a supine or prone position. None of the volunteers reported discomfort, and the helmet was generally perceived as non-restrictive when used in a seated posture. Potentially, our ultrasound imager can accommodate a range of functional tasks for different brain regions, including finger tapping, tongue tapping, language tasks, and even decision-making tasks in the future. Overall, the imager enables both brain anatomy imaging and functional studies at a high spatiotemporal resolution. It can be adjusted to target different brain regions based on the location of the acoustic window. Furthermore, it offers potential applications in brain functional tasks, including those involving free movement or brain-machine interfaces[31].

This study focused on demonstrating the feasibility of helmet-based ultrasound imaging in post-hemicraniectomy patients, a platform with several promising clinical applications. One potential future use is the continuous monitoring of brain function in both seated and free-moving conditions.

For example, the current system designed for post-hemicraniectomy patients could be adapted for routine monitoring, such as assessing scalp blood flow to evaluate healing and tracking functional recovery over time. This capability could enable early detection of brain dysfunction and provide insight into patients' neurological status during daily life, offering valuable support for rehabilitation and long-term care. Beyond monitoring, the system may also contribute to the development of brain-machine interfaces (BMI)[35]. By decoding brain signals noninvasively, the technology could potentially support BMI-based control of assistive devices, such as mobile chairs, improving patient autonomy. Furthermore, the integration of imaging and stimulation in a closed-loop system could open new avenues for neuromodulation research—leveraging ultrasound or other external stimulators to assess the brain's functional response in real time. Because the helmet ultrasound system enables high-resolution imaging of brain soft tissue and vasculature, it may also support long-term monitoring of brain tumors or cerebrovascular conditions such as stroke[36]. For select patients, the clinical risk of undergoing craniectomy to enable continuous brain imaging or BMI access may be justifiable, particularly when compared to the challenges of traditional invasive interfaces that require electrode implantation and suffer from long-term signal degradation[37]. In such cases, a craniectomy window could offer a stable platform for chronic, noninvasive monitoring and interface deployment without the need for repeated surgical interventions. Additionally, our system is modular and noninvasive, supporting flexible expansion to larger cohorts and additional imaging modes, including its potential future application to the patient cohort with an acoustically transparent cranial window[38].

The helmet-based ultrasound system has the potential to complement traditional imaging modalities such as transcranial ultrasound and X-ray CT for brain imaging. Conventional transcranial ultrasound is typically limited to visualizing large basal cerebral arteries through acoustic windows like the temporal bone and lacks the resolution and sensitivity needed to detect cortical vasculature in adult patients, primarily due to skull-induced attenuation and distortion. In contrast, our approach takes advantage of the skull-less window in post-hemicraniectomy patients, allowing direct, high-resolution imaging of both cortical and subcortical structures, including small blood vessels. Furthermore, the two approaches can be combined using a helmet-mounted transcranial ultrasound system with wide applicability in the general adult population. While CT offers detailed anatomical information, vascular imaging typically requires contrast agents, limiting its use for routine or continuous monitoring. Additionally, the exposure to ionizing radiation makes CT less suitable for repeated imaging, especially in neurocritical care settings. Nevertheless, based on the high contrast of these features in the brain in Fig. 3, we can co-register the brain structure images obtained by the helmet-based ultrasound scanner with established brain imaging techniques, such as X-ray CT and MRI, in the future. A comparison of X-ray CT, MRI, fNIRS, conventional transcranial ultrasound, and our work is provided in Supplementary Table 1.

Although the current system enables hands-free imaging with a helmet-based probe, it remains tethered to a benchtop ultrasound platform (Verasonics), limiting its portability and practical use in ambulatory or home-care settings. This trade-off was necessary to establish and validate core imaging capabilities in awake human subjects, including 3D structural and blood flow visualization. We acknowledge that this setup does not yet constitute a fully wearable system. However, the modular helmet design allows for future integration with portable, low-power ultrasound electronics, including commercially available smartphone-based ultrasound probes[39]. Ongoing efforts are focused on miniaturizing the acquisition hardware to advance toward a compact, self-contained imaging unit suitable for continuous or mobile brain monitoring in clinical and non-clinical environments.

This work signifies a step towards demonstrating the feasibility of utilizing ultrasound as a wearable human brain imaging technology. It focuses on developing a helmet-based wearable ultrasound brain imager that provides 3D imaging of deep human brain tissue structures, vasculature, and blood flow with high spatiotemporal resolutions, including the brain cortex regions. The technology holds promise as a wearable brain imaging tool, particularly in post-hemicraniectomy patients. This unique patient population offers valuable opportunities for brain imaging, potentially providing insights that surpass those obtained from non-human primate studies. Previous work, however, has either focused on bedside functional ultrasound imaging of the neonatal brain[23] or transcranial imaging of major cerebral arteries with a conformal ultrasound patch due to the small acoustic window in the human head[30]. Future endeavors should prioritize several key areas of development: 1) Advancing the device's design to make it lighter and more compact, potentially exploring the use of an ultrasound patch[26,30]. 2) Expanding the scope of performance evaluation to a broader population, including neonates and healthy adults. 3) Adjusting the imager for transcranial imaging applications using lower center frequency probes attached to the natural acoustic window (i.e., temporal bone) of the head. 4) Extending its applicability to various brain functions and capabilities, such as motor function, language function, working memory, and decision-making[35]. 5) Exploring potential clinical applications, such as aiding in the monitoring and diagnosis of brain tumors or strokes[36]. 6) Exploring the potential of integrating it into brain-machine interfaces, enabling interaction with the external environment[31,37]. 7) Investigating the synergistic benefits of brain stimulation to a closed-loop system with this wearable device[40].

This study also marks an initial step toward developing a technically feasible and adaptable wearable ultrasound platform for brain imaging. The proposed helmet-based system enables stable acoustic coupling and supports 3D volumetric imaging with high spatiotemporal resolution in post-hemicraniectomy patients. Its modular design allows compatibility with different transducer configurations and adaptation to various imaging needs. Future work will focus on reducing the system's size and weight, improving robustness to motion, and enhancing continuous and long-term monitoring capabilities. We also plan to explore integration with complementary modalities, such as photoacoustic tomography, and investigate potential applications in closed-loop neuromodulation. These developments will further support the system's translation to applications such as functional imaging, neurological monitoring, and brain–machine interfaces.

**Materials and Methods**

**Helmet ultrasound imager construction**

The helmet ultrasound imager consists of a customized helmet onto which a motorized linear translation stage (LCR30, Parker; travel length: 75 mm, bidirectional repeatability: $\pm 0.1$ mm) and a customized ultrasound transducer array (center frequency: 5 MHz, 128 elements, element width: 0.25 mm, element pitch: 0.298 mm) are mounted. The customized helmet is made by modifying a helmet to include the imaging window and mounting screw-holes for the motorized linear translation stage. We use custom-made 3D-printed parts to attach the motor to the helmet and the probe to the motor. Four parts attach the probe to the motor, as shown in Supplementary Fig. 3— the motor attachment, the probe holder, and two flexible mounts (FM1 and FM2). FM1 can rotate about the motor attachment, and FM2 can rotate about and translate along FM1. This four-part design allows adequate flexibility to position the probe so it can conform to the shape of the head. The modular design also makes it easy to switch to a different transducer array, simply by designing an appropriate probe holder.

The ultrasound probe was connected to a Verasonics Vantage system (Verasonics Inc.), which featured 14-bit A/D converters, a 62.5 MHz sampling rate, and a programmable gain of up to 51 dB. The acquired ultrasound signals were digitized and stored within the Verasonics system and subsequently streamed to a computer via PCI Express. To program the Verasonics system, a custom MATLAB script was run on this computer. The motorized linear translation stage was controlled by a dedicated MATLAB application, which communicated with an Arduino Mega board (Arduino, Somerville, MA, USA) through a USB 3.0 connection. The Arduino board, in turn, controlled a stepper motor driver, which drove the motor. A schematic illustrating the data acquisition and control module is provided in Supplementary Fig. 8.

**Imaging sequence and data processing**

For all the structure ultrasound images in Figs. 2 - 4, we acquired a compounded frame using 15 plane waves[41] at equally spaced angles between $\pm 14°$ using a pulse repetition frequency of 8000 Hz. For all the functional ultrasound images in Figs. 2 - 4, we acquired 300 compounded frames at a compounded frame rate of 300 Hz to 500 Hz, where each compounded frame was acquired using the same parameters as stated above. The 300 frames were then processed using a singular value decomposition (SVD) clutter filter[37] as shown in Supplementary Fig. 9, which separates the blood signals from the tissue signals, to generate the fUS images of blood flow. The effect of the singular value cutoff of the SVD clutter filter on a representative *in vivo* blood flow image is shown in Supplementary Fig. 10. To acquire 3D volumes, the motorized linear stage was used to move the probe over a range of 5 cm, in steps of 2 mm. At each step, a 2D fUS image was acquired using the above-stated parameters, and the acquired slices were stitched together to create the 3D volumetric images. The motor and the Verasonics system were controlled using a custom Python script running on the computer. The mechanical index for the ultrasound output was 0.8 (<1.9).

For the results in Fig. 5, each compounded frame was acquired using 5 plane waves at equally spaced angles between $\pm 6°$ using a pulse repetition frequency of 6000 Hz. 120 compounded frames were acquired at a frame rate of 300 Hz, and an SVD clutter filter was applied. fUS images were acquired every 2.5 seconds over a duration of 120 seconds. The inter-frame motion, modeled as a rigid transformation, was corrected using intensity-based automatic image registration tools in the MATLAB image processing toolbox, with mean square error as the optimization metric and bilinear interpolation as the image resampling method[43].

**Experimental setup**

To construct the line-target phantom, we 3D-printed a U-shaped structure with holes on either side. We ran a wire of diameter 150 $\mu$m through the holes to create a grid of 4×3 parallel lines. We then placed the phantom in a water tank and immersed the ultrasound array in water to ensure acoustic coupling. To construct the blood flow phantom, we used the same 3D printed structure, but with the blood tube passing through the holes arbitrarily. We used micro-renathane tubing (MRE080-S3850, Braintree scientific) with inner and outer diameters of 0.97 mm and 2.03 mm, respectively. We then perfused 45% hematocrit whole bovine blood (QuadFive) through the tubing. Schematics of the line-target and blood flow phantoms are shown in Supplementary Fig. 4.

For the *in vivo* experiments, we instructed the hemicraniectomy subject to wear the helmet and sit in a chair. To facilitate the coupling of ultrasound from the probe to the head, we used an agarose gel pad attached to the probe. This minimized the formation of air bubbles compared to using large quantities of ultrasound gel, while remaining compressible enough to conform to different head shapes. During the motion of the probe, the gel pad was attached to the probe with a plastic film and moved together with it. To ensure hygiene and minimize contamination risk during imaging, both pre- and post-scan cleaning protocols were followed. Before each scan, the subject's scalp area was gently cleaned using alcohol wipes, and a fresh, sterile ultrasound gel pad was applied. After imaging, the gel residue was removed using disposable towels, and the scalp was cleaned again with alcohol wipes. Reusable components in contact with the skin, such as the helmet interface and transducer holders, were disinfected using medical-grade cleaning agents between sessions. These measures ensured a consistent standard of cleanliness and safety for all participants.

**Imaging protocols.** The experiments on human imaging were performed in a dedicated imaging room. All experiments were performed at Caltech according to the relevant guidelines and regulations approved by the Institutional Review Boards of the California Institute of Technology (Caltech), the University of Southern California (USC), and the Rancho Los Amigos National Rehabilitation Center (RLA). Four adult hemicraniectomy patients (participant 1: left hemicraniectomy, male; participant 2: right hemicraniectomy, male; participant 3: right hemicraniectomy, female; participant 4: left hemicraniectomy, male) with completely healed surgical wounds were recruited from RLA. Written informed consent was obtained from all the participants according to the study protocols.

**Data availability**

The data that support the findings of this study are provided within the paper and its supplementary material.

**Code availability**

The reconstruction algorithm and data processing methods can be found in the Methods. The reconstruction code is not publicly available because it is proprietary and is used in licensed technologies.

**Acknowledgments**

This work was sponsored by the United States National Institutes of Health (NIH) grants U01 EB029823 (BRAIN Initiative) and R35 CA220436 (Outstanding Investigator Award).

**Contributions**

L.V.W., C.Y. L., J.J.R., and Y.Z. designed the study. C.Y.L. and J.J.R. recruited the subjects. Y.Z., K.S., I.R., and J.O.G. designed and built the system. Y.Z., K.S., J.J.R., I.R., and J.O.G. performed the experiments. Y.Z. and K.S. analyzed the data and wrote the manuscript with input from all of the authors. L.V.W., C.Y.L., and J.J.R. supervised the study and revised the manuscript.

**Competing interests**

L.V.W. has a financial interest in Microphotoacoustics Inc., CalPACT LLC, and Union Photoacoustic Technologies Ltd., which, however, did not support this work.

# Supplementary Materials

## Supplementary Table

Supplementary Table 1 | Comparison of X-ray computed tomography (CT), magnetic resonance imaging (MRI), functional near-infrared spectroscopy (fNIRS), conventional transcranial ultrasound, and this work

## Supplementary Figures

Supplementary Fig. 1 | Photographs of the helmet ultrasound imager.
Supplementary Fig. 2 | Illustrations of the ultrasound imaging sequences.
Supplementary Fig. 3 | Custom-designed 3D-printed parts on the helmet allow for flexibility in the positioning of the ultrasound array.
Supplementary Fig. 4 | Phantom design.
Supplementary Fig. 5 | Quantification of the diameters of fine vessels.
Supplementary Fig. 6 | Long-term monitoring of brain vasculature.
Supplementary Fig. 7 | Brain tissue structure, power Doppler, and color Doppler images.
Supplementary Fig. 8 | Electrical schematic of the wearable ultrasound system.
Supplementary Fig. 9 | Spatiotemporal filtering schematic.
Supplementary Fig. 10 | Choice of the SVD threshold.

**Supplementary Table**

Supplementary Table 1 | Comparison of X-ray computed tomography (CT), magnetic resonance imaging (MRI), functional near-infrared spectroscopy (fNIRS), conventional transcranial ultrasound, and this work

| Modality | Imaging type | Operator dependence | Ionizing radiation | Portability | Spatial resolution | Temporal resolution | Bedside applications | Wearable capabilities |
|---|---|---|---|---|---|---|---|---|
| X-ray CT | Structural | Low | Yes | No | Moderate (sub-mm to mm) | Low (seconds) | No | No |
| MRI | Structural & functional | Low | No | No | Moderate (sub-mm to mm) | Low (seconds-minutes) | No | No |
| fNIRS | Functional (hemodynamics) | Medium | No | Yes | Low (cm) | High (ms-s) | Yes | Yes |
| Conventional transcranial ultrasound | Structural & flow (major arteries) | High | No | Yes | Moderate (mm) | High (ms) | Yes | Feasible |
| Helmet ultrasound (This Work) | Structural & flow | Low (Hands-free design) | No | Yes | High (sub-mm) | High (ms) | Yes | Yes |

**Supplementary figures**

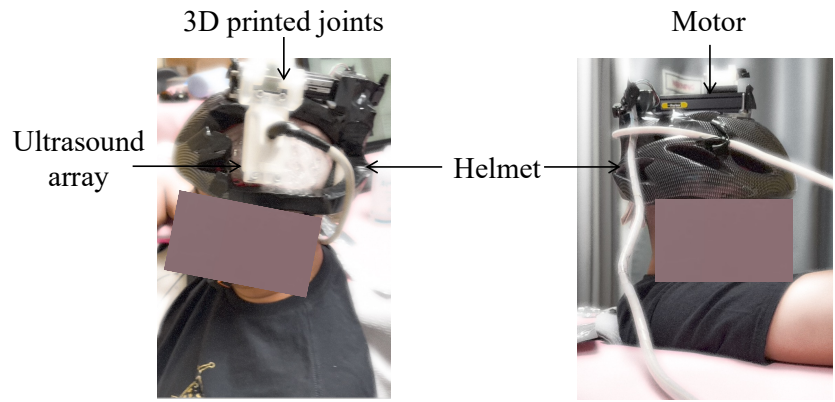

**Supplementary Fig. 1 | Photographs of the helmet ultrasound imager.** It comprises four integrated components: an ultrasound array for transmitting ultrasound pulses and detecting backscattered ultrasound waves from the head, a motorized linear translation stage for scanning the array to form a three-dimensional (3D) field-of-view, a customized helmet to support the ultrasound array with the motor and ergonomically fit on the human head, and custom-made 3D-printed parts to provide flexibility in adjusting the position and orientation of the ultrasound array.

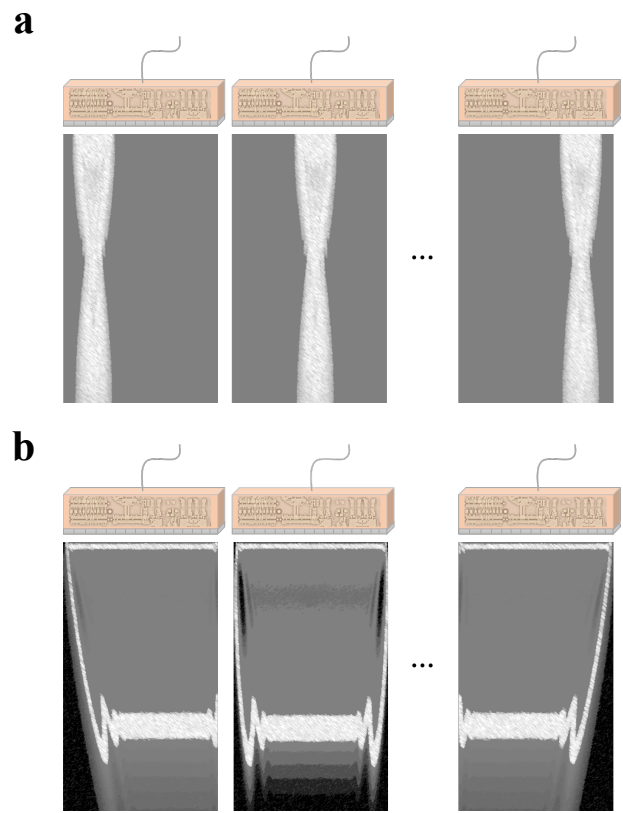

**Supplementary Fig. 2 | Illustrations of the ultrasound imaging sequences. a,** Focused beam imaging sequence**; b,** Tilted plane wave imaging sequence.

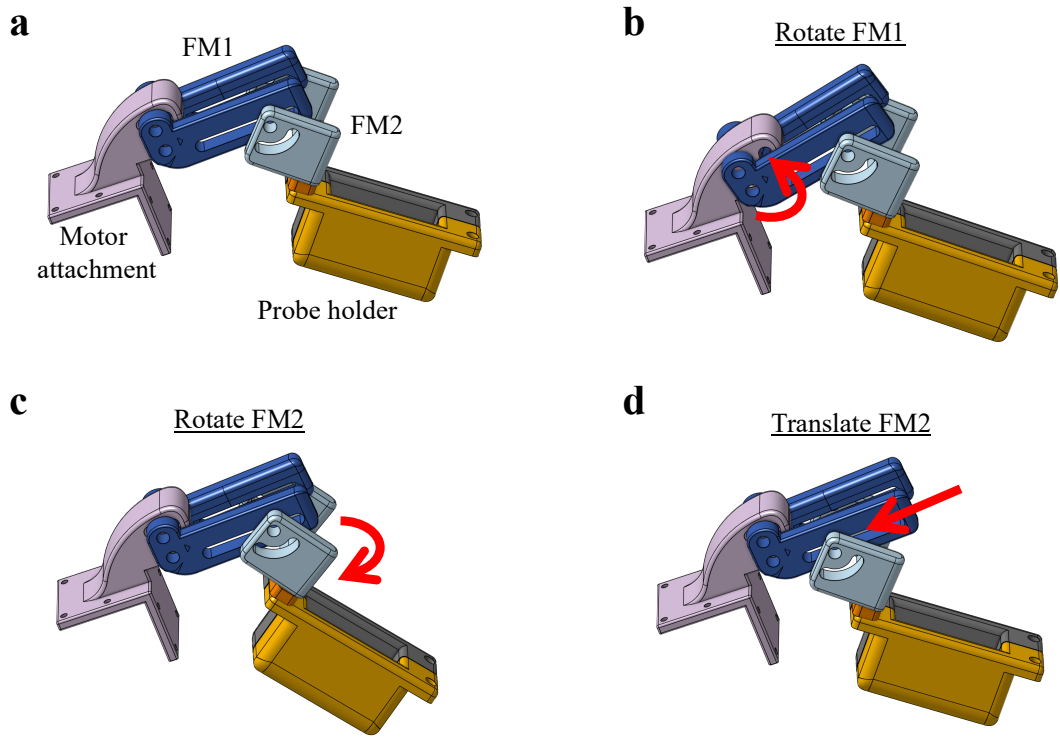

**Supplementary Fig. 3 | Custom-designed 3D-printed parts on the helmet allow for flexibility in the positioning of the ultrasound array. a,** Four parts attach the probe to the motor: the motor attachment, the probe holder, and two flexible mounts, FM1 and FM2. This four-part design provides adequate flexibility to position the probe at various angles and locations, depending on the shape of the head. **b–d,** Illustrations of three configurations of the designed parts, accomplished by rotating FM1 about the motor attachment, rotating FM2 about FM1, and translating FM2 along FM1, respectively.

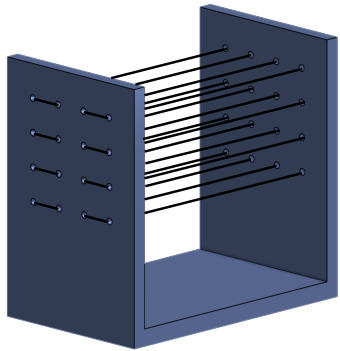 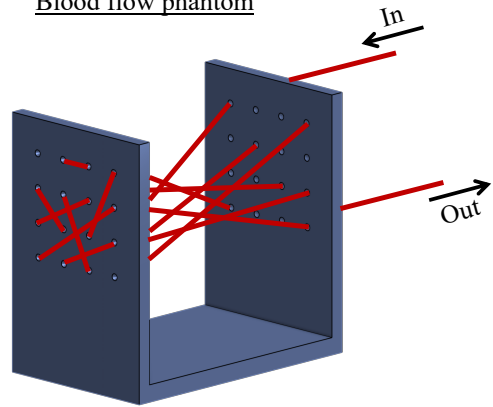

**Supplementary Fig. 4 | Phantom design. a,** Illustration of the line-target phantom. A U-shaped frame with a grid of holes on either side is 3D-printed. A wire of 0.15 mm diameter is run through the holes to create a 4×3 grid of parallel wires. **b,** Illustration of the blood flow phantom. Micro-renathane tubing (MRE080-S3850, Braintree scientific; outer diameter: 2.03 mm, inner diameter: 0.97 mm) is run through the same U-shaped frame to create an arbitrary pattern of tubing. The tubing is perfused with 45% hematocrit whole bovine blood (QuadFive).

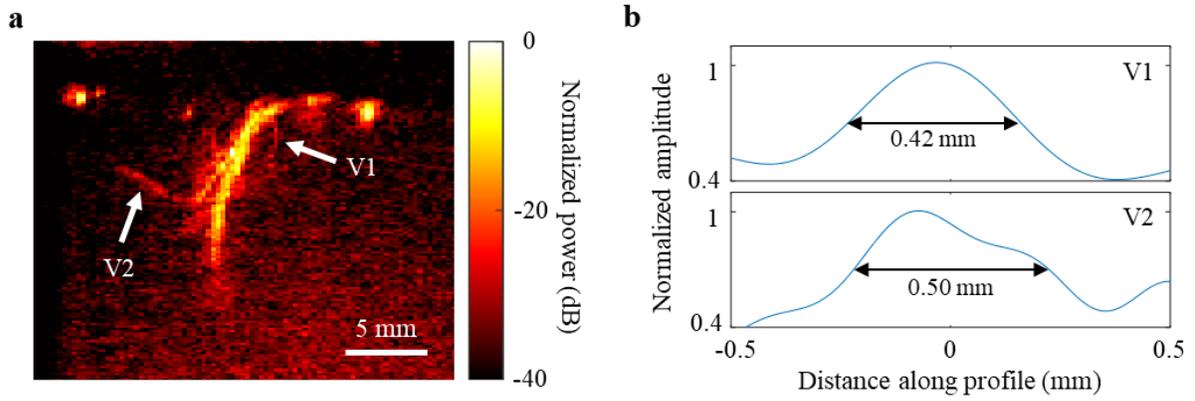

**Supplementary Fig. 5 | Quantification of the diameters of fine vessels. a,** 2D blood vasculature image acquired from a hemicraniectomy subject. Two fine vessels, V1 and V2, are indicated by white arrows. **b,** Profiles of the vessels V1 and V2. The diameters of the two vessels, computed as the full width at half maximum above the background (0.4 times the maximum of the profile), are 0.42 mm and 0.5 mm, respectively.

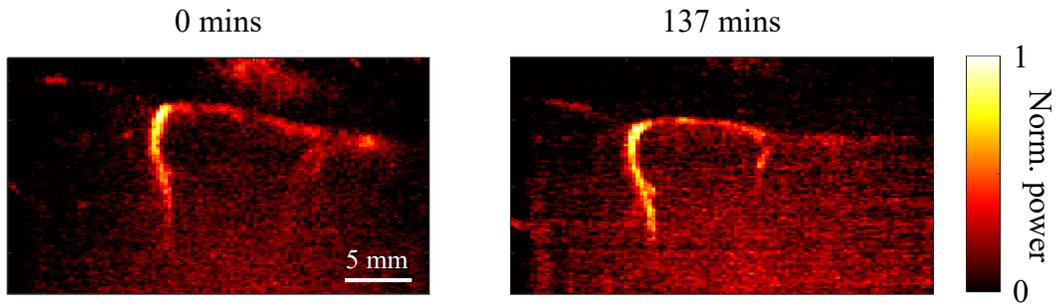

**Supplementary Fig. 6 | Long-term monitoring of brain vasculature.** 2D blood flow images of the same cross-section of the brain in a hemicraniectomy subject (in a sitting position), obtained 137 minutes apart. The helmet was removed after the first image acquisition and then repositioned on the subject's head approximately two hours later. Despite the 137-minute interval, similar brain vasculature features are observed in both images, with minor discrepancies attributed to a slight shift in the probe's imaging position.

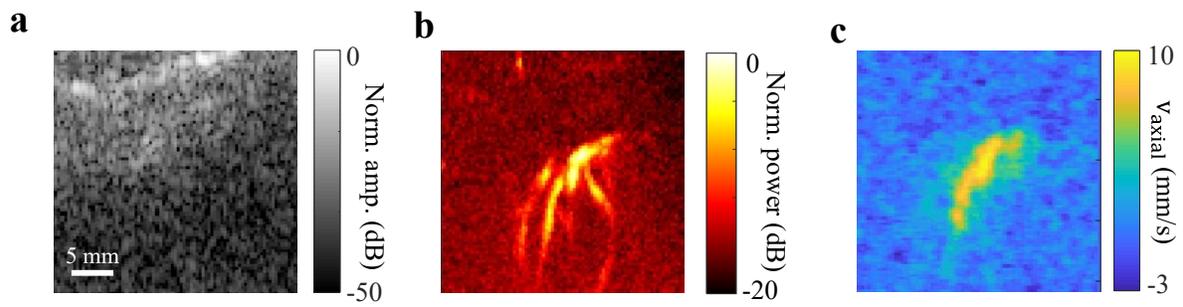

**Supplementary Fig. 7 | Brain tissue structure, power Doppler, and color Doppler images. a-c,** 2D ultrasound cross-section, power Doppler image, and color Doppler image, respectively, of the human brain. $v_{axial}$ represents the axial component (relative to the probe surface) of the blood flow velocity.

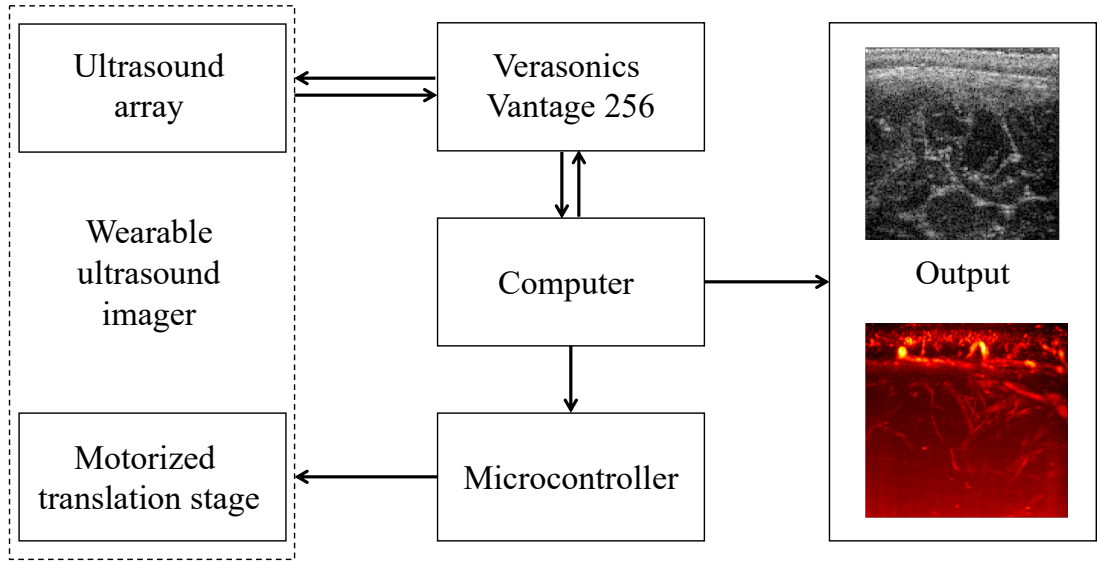

**Supplementary Fig. 8 | Electrical schematic of the helmet-based wearable ultrasound imager.** The ultrasound array is connected to a Verasonics Vantage 256 system (Verasonics Inc.), which controls the transmission from the ultrasound array and acquires, digitizes, and stores the backscattered signals. The data are streamed to a computer via PCI Express. The computer is also used to program the Verasonics system using a custom MATLAB script. A MATLAB application on the computer controls the motorized linear translation stage via an Arduino Mega (Arduino, Somerville, MA, USA) microcontroller. Finally, the computer is used to reconstruct the ultrasound structure and fUS images of the brain using the acquired ultrasound signals.

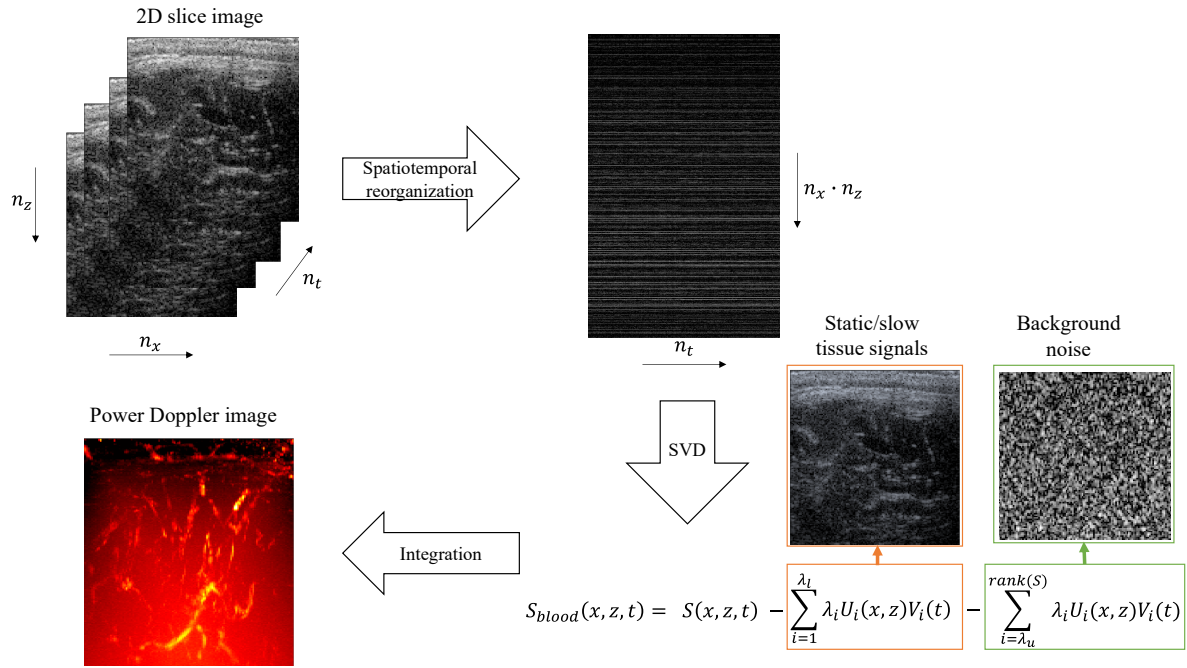

**Supplementary Fig. 9 | Spatiotemporal filtering schematic.** A 2D image stack ($n_z \times n_x$) of structural ultrasound images are acquired across time, forming a 3D spatiotemporal matrix with dimensions $n_z \times n_x \times n_t$ (top left). The 3D matrix is reshaped to 2D with dimensions $n_x n_z \times n_t$ (top right). Singular value decomposition is performed on the reshaped matrix, after which the blood signal is extracted by removing the slow-moving tissue and background noise corresponding to the small ($\lambda_l$) and large ($\lambda_u$) singular value indices, respectively. Power Doppler images are generated from the integration of the blood signal. SVD: Singular value decomposition; 2D: two-dimensional.

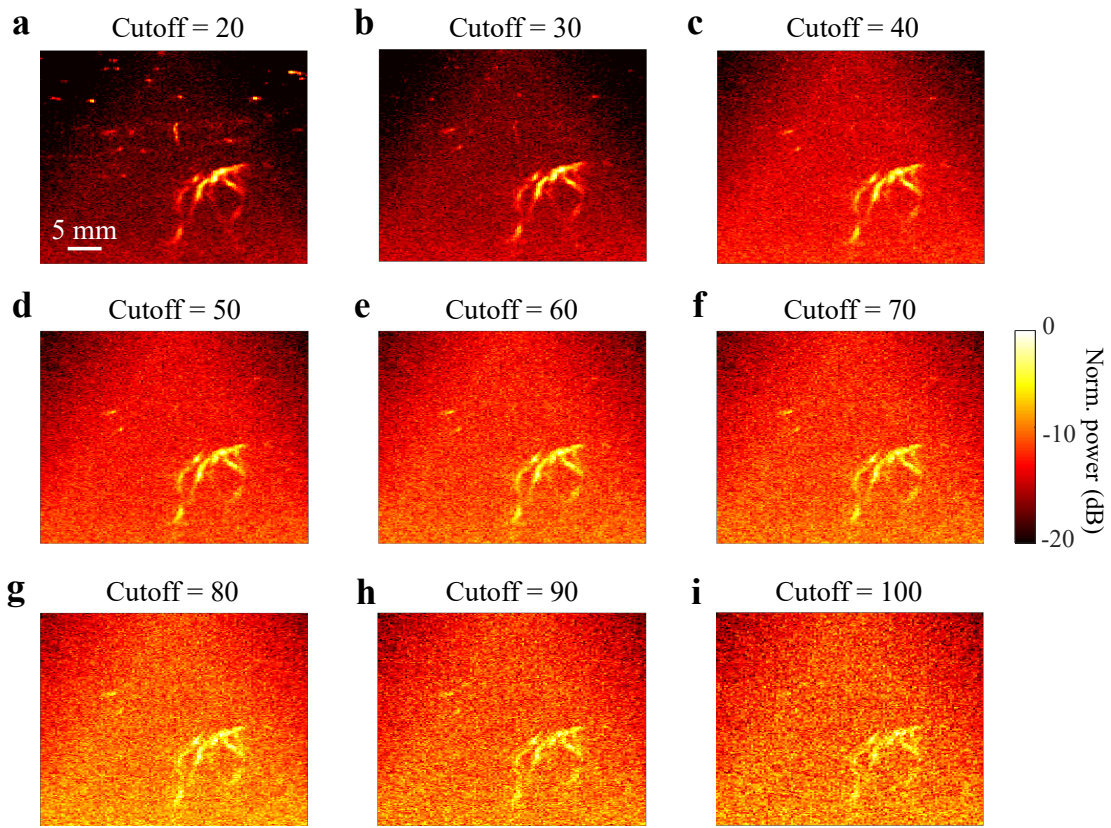

**Supplementary Fig. 10 | Choice of the SVD threshold. a-i,** fUS images showing brain vasculature obtained after power Doppler processing using different thresholds for the SVD clutter filter.